\documentclass[conference]{IEEEtran}
\IEEEoverridecommandlockouts
\usepackage{cite}
\usepackage{amsmath,amssymb,amsfonts}
\usepackage{algorithm}
\usepackage{algorithmic}
\usepackage{graphicx}
\usepackage{textcomp}
\usepackage{xcolor}
\usepackage{amsthm}
\theoremstyle{definition}
\newtheorem{definition}{Definition}[section]

\usepackage{multirow}

\def\BibTeX{{\rm B\kern-.05em{\sc i\kern-.025em b}\kern-.08em
    T\kern-.1667em\lower.7ex\hbox{E}\kern-.125emX}}
\begin{document}



\title{Developing An Attention-Based Ensemble Learning Framework for Financial Portfolio Optimisation}



\author{\IEEEauthorblockN{Zhenglong Li}
\IEEEauthorblockA{
\textit{Department of Electrical and Electronic Engineering} \\
\textit{The University of Hong Kong}\\
Hong Kong SAR, China \\
lzlong@hku.hk}
\and
\IEEEauthorblockN{Vincent Tam}
\IEEEauthorblockA{
\textit{Department of Electrical and Electronic Engineering} \\
\textit{The University of Hong Kong}\\
Hong Kong SAR, China \\
vtam@eee.hku.hk}
}

\maketitle

\begin{abstract}
In recent years, deep or reinforcement learning approaches have been applied to optimise investment portfolios through learning the spatial and temporal information under the dynamic financial market. Yet in most cases, the existing approaches may produce biased trading signals based on the conventional price data due to a lot of market noises, which possibly fails to balance the investment returns and risks. Accordingly, a multi-agent and self-adaptive portfolio optimisation framework integrated with attention mechanisms and time series, namely the MASAAT, is proposed in this work in which multiple trading agents are created to observe and analyse the price series and directional change data that recognises the significant changes of asset prices at different levels of granularity for enhancing the signal-to-noise ratio of price series. Afterwards, by reconstructing the tokens of financial data in a sequence, the attention-based cross-sectional analysis module and temporal analysis module of each agent can effectively capture the correlations between assets and the dependencies between time points. Besides, a portfolio generator is integrated into the proposed framework to fuse the spatial-temporal information and then summarise the portfolios suggested by all trading agents to produce a newly ensemble portfolio for reducing biased trading actions and balancing the overall returns and risks. The experimental results clearly demonstrate that the MASAAT framework achieves impressive enhancement when compared with many well-known portfolio optimsation approaches on three challenging data sets of DJIA, S\&P 500 and CSI 300. More importantly, our proposal has potential strengths in many possible applications for future study.

\end{abstract}


\section{INTRODUCTION}

Portfolio management (PM) is a crucial aspect of investment decision making, with a primary objective of achieving higher returns while effectively reducing risks through dynamically allocating capital to assets in a portfolio. Due to the highly volatile financial markets where the movements of asset prices are driven by various factors such as the global economy, company operation, and public sentiment, it is still a challenge to manage an optimal portfolio fulfilling the two opposite objectives. Conventionally, lots of financial models have been deployed in real-world markets in terms of varied investment principles like follow-the-loser~\cite{li2012pamr}, follow-the-winner~\cite{helmbold1998line}, and pattern matching~\cite{gyorfi2006nonparametric}. However, these works are typically effective in a single market condition and may easily fail in multiple trading periods with complex market dynamics. Recently, there is an increasing emphasis on the utilisation of machine learning techniques~\cite{gunjan2023brief} to extract the underlying patterns from the non-stationary price series in PM. Among these approaches, deep learning (DL) or reinforcement learning (RL)-based trading strategies~\cite{hambly2023recent} have achieved significant progress in the field of computational finance through developing trading agents to quickly learn the rewards of the newly generated portfolio and accordingly adjust trading strategies under the turbulent markets. Yet in most cases in the financial market, the price data is intrinsically a kind of time series that may involve a lot of noise for which the useful information for implying future trends can be difficult to recognise.

To better predict the trends of asset prices and also the dependencies between assets, more efforts~\cite{jiang2021applications} have been made to apply recurrent-based models for analysing price series. Nevertheless, these recurrent-based models are not performing well in capturing the correlations between stocks in PM for diversifying investment risks in a turbulent financial market. Meanwhile, handling long sequences in recurrent-based models can be very challenging due to the issues of gradient exploding and vanishing during the model training period. Alternatively, there are some interesting studies~\cite{chen2021novel} utilising convolution kernels to extract the correlations between assets through grouping the features of different assets as a 2-D feature map. For instance, the columns represent different assets while each row describes the features of an individual asset. Yet the resulting correlation features are highly sensitive to the relative positions of assets in input feature maps, the convolution-based models may partially capture the local correlations from the neighbouring assets in feature maps due to the size of kernels.

To address the above pitfalls, a \textbf{m}ulti-\textbf{a}gent and \textbf{s}elf-\textbf{a}daptive trading framework integrated with \textbf{a}ttention mechanisms and \textbf{t}ime series namely the \textbf{MASAAT} is proposed in this work in which multiple agents are created to observe and analyse the directional changes (DC) of asset prices in different levels of granularity so as to carefully revise the portfolios for balancing the overall returns and investment risks in a highly volatile financial market. Through the predefined DC filters using different thresholds to record the significant price changes, the agents firstly extract the DC features from conventional time-based price series, trying to be keenly aware of the transitions of market states under multiple viewpoints. Similar to the words in language processing, patches in image processing, and frames in speech recognition, this work provides a new way to generate the tokens in a sequence such that the attention-based cross-sectional analysis (CSA) module and temporal analysis (TA) module of the agents created by the proposed framework can effectively capture the correlations between assets and the dependencies between time points from DC features or price series. More specifically, by reconstructing the feature maps, the token of sequences in the CSA module is based on the features of individual assets, targeting to optimise the attention score embeddings between assets while the token of sequences in the TA module is based on the features of individual time points, attempting to highlight the relevance between current and previous time points. Furthermore, the dependencies information of assets and time points are fused in the spatial-temporal fusion block. Through the clear division of works between the CSA and TA modules to continuously analyse the trend patterns of assets in a portfolio, the agents will get more insights to suggest portfolios in terms of their specific viewpoints. Ultimately, the provision of the suggested portfolios by different agents is merged as a newly ensemble portfolio to quickly respond to the financial market. Even any particular agent fails to estimate market trends and generates biased suggestions, the proposed MASAAT framework integrated with multiple agents can still adaptively rectify the final portfolio to reduce the adverse effects. During the model training, the PM problem can be formulated as a partially observable Markov decision process (POMDP) and then optimised by a policy gradient method to carefully balance the trade-off between the long-term profits and risks in multiple trading periods. It should be noted that the newly proposed MASAAT is drastically different from a previous work~\cite{limasa2024} on a multi-agent framework for financial portfolio optimisation in which the DC and sequences of encoders of the transformer network are utilised to adapt to the time-varying financial markets. The main contributions of the proposed framework are summarised as follows.

\begin{enumerate}
\item Compared with the existing frameworks solely learning from conventional price series, our proposal utilises the DC features to capture the significant changes of price data in different levels of granularity, which can effectively enhance the signal-to-noise ratio of financial data and observe the dynamics of asset price from multi-scale receptive fields for further analysis.
\item The MASAAT framework provides a new way to generate the token of sequences for financial data such that the self-attention mechanism of the proposed CSA and TA modules in each agent can accordingly capture correlations between assets and the relevance of time points. Besides, the cooperating agents based on different viewpoints can reduce biased trading actions in the ever-changing financial market.
\item The attained empirical results on three challenging data sets reveal the potential benefits of our proposal integrated with multiple adaptive agents against other well-known PM methods in highly volatile financial markets.
\end{enumerate}

\section{RELATED WORKS}
\subsection{Directional Change}

In a highly volatile financial market, investors always concern about when and how the market style changes such that they can adjust their own portfolios to adapt to the fluctuating market dynamics. DC is an alternative event-based approach to sample data from the original time-based price series, which recognises the turning point of trends as a significant event to represent the current state of assets in a certain period. Compared with the traditional way to record the price data with fixed time intervals, the DC-based trading methods~\cite{ao2019trading} improve data efficiency and have achieved success in foreign exchange markets. From the perspective of DC, the movements of asset prices can be summarised into upward and downward trends. Each trend further consists of a DC event and an overshoot (OS) event. More specifically, given a pre-defined DC threshold $\Delta x_{dc}$, an upward DC event is confirmed when $p_{t} \ge p^{l}_{t-1} \times (1+\Delta x_{dc}) $, where $p_{t}$ is the current asset price, and $p^{l}_{t-1}$ is the lowest price in the last downward trend. Then the state turns into a period of upward OS events until the next downward DC event takes place. Conversely, a downward DC event is recognised when $ p_t\le p^{h}_{t-1} \times (1-\Delta x_{dc}) $, where $p^{h}_{t-1}$ is the highest price in the last upward trend. Followed by the confirmation of downward DC events, the downward OS event will start until the next upward DC event occurs. Yet the previous DC-based approaches with the use of meta-heuristic algorithms such as genetic programming~\cite{gypteau2015generating} and genetic algorithm~\cite{kampouridis2017evolving} did not carefully investigate the correlations between assets and the dependencies between time points in PM. This may not timely adjust portfolios for diversifying risks, especially when the market is highly turbulent.

\subsection{Attention-based Mechanism}

In recent years, the attention-based mechanism is a very promising research direction of artificial intelligence~\cite{chaudhari2021attentive, brauwers2021general} in which the frameworks based on different attention variants have been successfully applied in many tasks like natural language processing~\cite{galassi2020attention}, computer vision~\cite{liu2021swin}, and speech recognition~\cite{dong2018speech}. Learning the relative correlations between variables through imitating the cognitive behaviour of human beings to concentrate on the regions of interest can promote neural networks finding the most important information to help make decisions from historical experience. Among these attention-based schemes, a very successful variant is the self-attention mechanism~\cite{vaswani2017attention} as the core component of the well-known language models like ChatGPT~\cite{openai2023gpt4} and LLaMA~\cite{touvron2023llama}. The potential benefit of the self-attention mechanism is that the optimisation of feature embeddings will be guided to capture the internal correlation of tokens in a sequence or the mapping of tokens between two sequences. Typically, there are three inputs called query, key, and value vectors in the self-attention mechanism in which the relevance of tokens between query vectors and key vectors is firstly calculated as the weighting to highlight the important mappings of tokens between the target and source. Then the weighting is assigned to the value vector to ultimately generate the correlations of tokens in a sequence when all three mentioned vectors are the same sequence or build the mappings of tokens between two sequences when the query is different from key-value pairs. Our proposal tries to reconstruct the feature maps to generate specific tokens of financial data so as to capture spatial and temporal information by utilising the self-attention mechanism.

\subsection{Portfolio Optimisation}

As one of the fundamental research directions in computational finance, portfolio optimisation is still a challenging task due to the highly turbulent financial markets. There have been many conventional methods~\cite{li2014online} trying to follow the momentum of price changes for attaining sustained profits. Yet it is difficult for these methods to respond to the ever-changing markets with a newly adaptive trading strategy, especially when occurring unpredictable incidents such as the COVID-19 pandemic, trade war, oil price crash and local regional conflicts. Recently, DL and RL techniques were widely applied to learn underlying patterns from the volatile financial markets, together with various neural network architectures such as attention-based transformer~\cite{xu2021relation,gao2023stockformer}, graph-based neural network~\cite{yin2022graph}, and convolution-based neural network~\cite{jiang2017deep} for catching the features that may imply the trends of future movements in terms of price series. Furthermore, some approaches consider two different neural network architectures in a united framework to learn the temporal and spatial information, respectively. For instance, ~\cite{zhang2020cost} combines a recurrent-based sequential information net and a dilated causal convolution-based correlation information net to update portfolios. ~\cite{wang2021deeptrader} integrates a temporal convolution layer, spatial attention layer, and graph convolution layer to learn causal relationships between stocks. Besides, there are other interesting studies~\cite{liang2021adaptive} attempting to include sentiment information from financial news data in PM. In addition, ~\cite{ shen2019kelly} presents an ensemble learning framework based on a growth optimal portfolio strategy while ~\cite{ faridi2023portfolio} tries to use different machine learning methods and then combines their signals by a majority vote method. As aforementioned, these methods may not be easy to capture the underlying patterns from price series due to a lot of noise and then generate biased trading actions. To technically evaluate the performance of a financial portfolio, some essential financial concepts and performance indicators are introduced as below.
\begin{definition}
    \label{def:annualreturn} 
(Annualised Return) Considering the compound annual growth, the annualised return (AR) is 
    \begin{equation}
        AR=((\frac{C_{T}}{C_{1}})^{\frac{T_{yr}}{T}} - 1) \times 100\%
    \end{equation}
where $C_{1}$ and $C_{T}$ are the portfolio values at $t=1$ and $t=T$, $T_{yr}$ is the number of trading days per year, and $T$ is the number of trading days at a trading period.
\end{definition}

\begin{definition}
    \label{def:mdd} 
(Maximum Drawdown) Assume that $ t1<t2\le t$, the maximum drawdown (MDD) is defined as below to measure the maximum possible loss at a specific period.
    \begin{equation}
                MDD_{t} = \operatorname{Max} ( \frac{C_{t1}-C_{t2}}{C_{t1}})
    \end{equation} 
\end{definition}

\begin{definition}
    \label{def:sharperatio} 
(Sharpe Ratio) The Sharpe Ratio (SR), as a popular and comprehensive performance indicator to consider both profits and risks, is denoted as 
    \begin{equation}
        \text{SR} = \frac{AR-r_f}{\sigma_{p}}
    \end{equation}
\end{definition}
\noindent where $r_f$ is the risk-free rate to indicate the theoretical rate of returns received on zero-risk assets (usually refer to the yield of long-term treasury bond), $\sigma_{p}= \sqrt{\frac{T_{yr}}{T-1} \sum_{t=1}^{T}\left(r_{t}-\overline{r_{t}}\right)^{2}}$ as another risk metric describes the annualised volatility of the trading strategy, $r_{t}$ is the rate of daily returns at $t^{th}$ day, and $\overline{r_{t}}=\frac{1}{T} \sum_{t=1}^{T}r_{t}$ is the average of the rate of daily returns.
Besides, two regulations of trading are followed in this work including 1) long-only constraint (i.e., $ w_{i, t}\geq 0 $) and 2) capital budget constraint (i.e., $\sum_{i=1}^{N} w_{i, t}=1$). $w_{i,t}$ is the weight of $i^{th}$ asset in a portfolio at time $t$.

\section{METHODOLOGY}

\subsection{The Overview of the Framework}

\begin{figure*}[htbp] 
    \centering
    \includegraphics[width=0.9\linewidth]{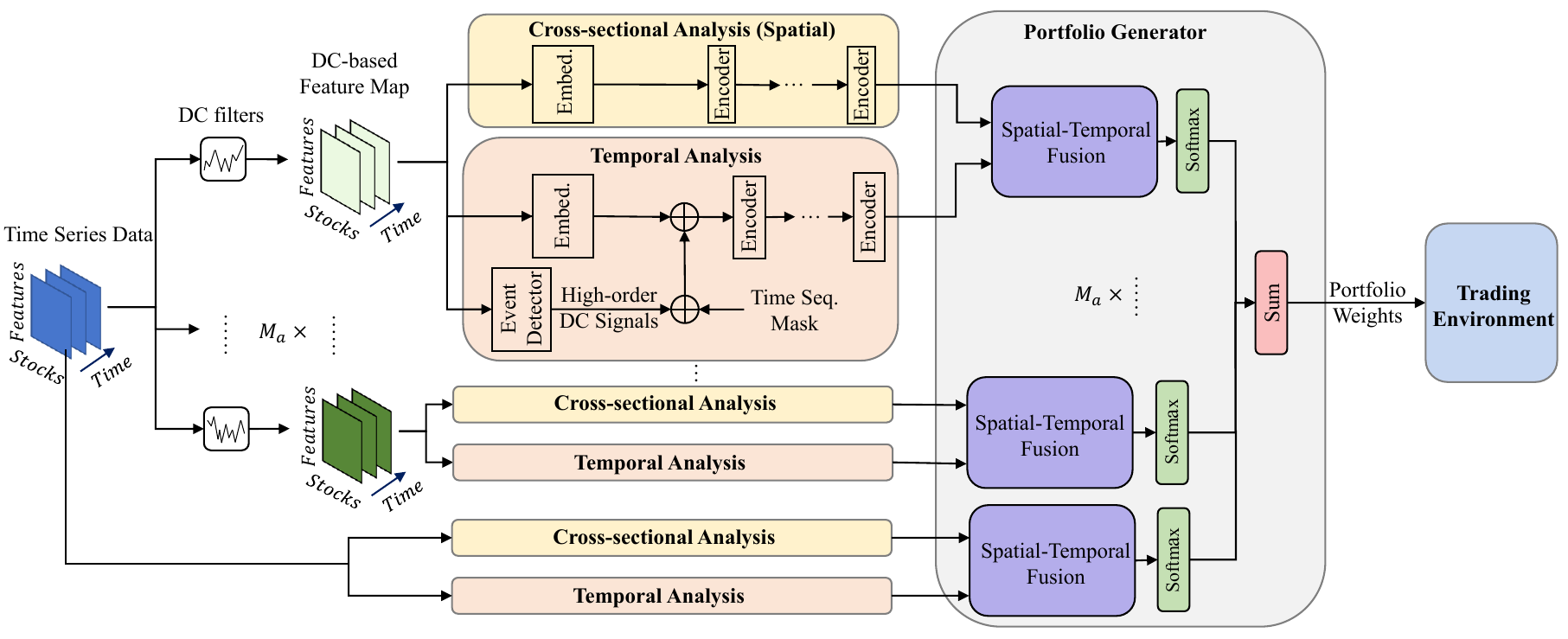}
    \caption{The System Architecture of the Proposed MASAAT Framework}
    \label{fig:overview}
\end{figure*}

\begin{algorithm}[tb]
    \caption{The Training Procedure of the MASAAT Framework}
    \label{algo:masaat} 
    \begin{algorithmic}[1]
        \STATE \textbf{Input}: $MaxIterations$ is the maximum iterations for training, $T_{m}$ is the number of trading days for a training mini batch, and $T_{w}$ is the observation period.
        \STATE \textbf{Output}: The best RL policy $\pi^{*}$. 
        \STATE Initialise the RL policy $\pi$, and the memory tuple $\hat{D}$.
        \FOR{$k=1$ to $MaxIterations$}
            \STATE Sample a day $t_{s}$ as the first trading day of the training mini batch.
            \FOR{$t=t_{s}$ to $t_{s}+T_{m}$}
                \STATE Observe the time series data $\mathbf{P}$ from $t-T_{w}+1$ to $t$.
                \STATE Invoke the multiple DC filters to generate DC feature maps $\mathbf{P_{DC}}$.
                \STATE Reconstruct the feature map $\mathbf{P_{DC}}$ to $\mathbf{P^{CSA}_{DC}}$ and $\mathbf{P^{TA}_{DC}}$.
                \STATE Reconstruct the feature map $\mathbf{P}$ to $\mathbf{P^{CSA}}$ and $\mathbf{P^{TA}}$.
                \STATE Invoke the CSA and TA modules to generate the embedding $\mathbf{O^{CSA}}$ and $\mathbf{O^{TA}}$ in each agent according to the RL policy $\pi$.
                \STATE Invoke spatial-temporal fusion module to fuse the $\mathbf{O^{CSA}}$ and $\mathbf{O^{TA}}$  in each agent.
                \STATE Integrate the trading signals from all agents to generate the new portfolio $\mathbf{w_{t}}$.
                \STATE Execute the new portfolio $\mathbf{w_{t}}$ and calculate the reward $r_{t}$.
                \STATE Store tuple ($w_{t}$, $r_{t}$, $\mathbf{P}$, $\mathbf{P_{DC}}$) in $\hat{D}$.

            \ENDFOR
            \IF {the policy update condition is triggered}
            \STATE Update the RL policy $\pi$ including the learnable parameters of CSA, TA, and fusion modules of each agent by learning the historical profile $\hat{D}$.
            \STATE Reset the memory tuple $\hat{D}$.
            \ENDIF

        \ENDFOR
        \STATE \textbf{return} the best RL policy $\pi^{*}$.
    \end{algorithmic}
\end{algorithm}

The system architecture of the proposed framework is shown in Fig.~\ref{fig:overview}. Compared with the previous studies on PM wherein the models learn price patterns from conventional price series, the proposed MASAAT framework applies multiple DC filters in terms of different DC thresholds to capture the significant changes of asset prices from multi-scale receptive fields for analysing the possible effects to future price movements. Specifically, the receptive fields in this work represent the different levels of asset price fluctuations, which provides multiple views for trading agents to intuitively perceive dynamic market states just as using different convolution kernel filters in computer vision to observe images in different scales. Furthermore, through reconstructing the asset-oriented DC features in the CSA module and time point-oriented DC features in the TA module as the tokens of sequences, the multi-agent scheme of the MASAAT framework may concurrently collect the spatial-aware and temporal-aware information in varied degrees of price changes to help deduce the direction and scale of future trends. Similarly, the original price series data will be directly reshaped to the asset-orient and time point-oriented price features as well, followed by the cross-sectional and temporal information extraction from the CSA and TA modules. It is worth noting that the CSA and TA modules are based on the encoders with self-attention mechanisms in which the attention scores are calculated by the global sequence on all tokens such that the similarity measurement of all assets can be fairly computed while the convolution-based neural network is very sensitive to the relative position of assets in feature maps and focuses on the local areas according to the convolution kernel size. On the other hand, due to the attention scores indicating the similarity of tokens, the trading signals generated by the proposed framework can be more explainable. Afterwards, by utilising the attention mechanism in the spatial-temporal block to build the mapping between the asset sequence and historical time point sequence, the trading agents generate the embeddings to represent the attention scores of each asset to every time point within the given observation window and then output their suggested portfolios, respectively. Lastly, the portfolio generator summarises all suggestions from different agents to produce a newly revised portfolio to adapt to the current financial environment.

As clearly illustrated in Algorithm~\ref{algo:masaat}, the pseudo-code of the training procedure of the proposed MASAAT framework is given. Assume that $N$ is the number of assets in a portfolio, $M$ is the number of observation features from financial markets, and $M_{a}$ is the number of trading agents. For the sampled trading days, the trading agent firstly observes the price features $\mathbf{P} \in \mathbb{R}^{N\times M \times T_{w}} $ within the observation period $T_{w}$. Then the DC-based features $\mathbf{P_{DC}}=\left \{ \mathbf{P_{DC,1}}, \mathbf{P_{DC, 2}},...,\mathbf{P_{DC, M_{a}}}    \right \} \in\mathbb{R}^{M_{a}}$, $\mathbf{P_{DC, i}} \in \mathbb{R}^{N\times M \times T_{w} }$ are obtained by DC filters. As aforementioned, the $\mathbf{P_{DC,i}}$ will be reconstructed to $\mathbf{P^{CSA}_{DC,i}} \in \mathbb{R}^{N\times M T_{w}}$ for the CSA module and $\mathbf{P^{TA}_{DC,i}} \in \mathbb{R}^{T_{w} \times NM }$ for the TA module to fit the transformer encoders with the fewest modifications. Similarly, the original price series $\mathbf{P}$ is converted to $\mathbf{P^{CSA}} \in \mathbb{R}^{N\times M T_{w}}$ and $\mathbf{P^{TA}} \in \mathbb{R}^{T_{w} \times NM }$. After measuring the similarities between tokens in the provided sequence, the CSA and TA modules output the asset-oriented embedding $\mathbf{O^{CSA}} \in \mathbb{R}^{N \times D}$  and time point-oriented embedding $\mathbf{O^{TA}} \in \mathbb{R}^{T_{w} \times D}$, where $D$ is the embedding size of a token. Subsequently, these embeddings are fused to generate a new portfolio and then further integrated with other agents to provide a final weight vector $\mathbf{w_{t}}$ for adjusting the portfolio. After executing the order, the reward $r_{t}$ will be collected and stored into the memory $\hat{D}$, together with $\mathbf{w_{t}}$, $\mathbf{P}$, $\mathbf{P_{DC}}$. Furthermore, the RL policy $\pi$ will be iteratively updated through learning the trading profile $\hat{D}$ with the use of the policy gradient method when the condition of the policy update is triggered. Ultimately, the best RL policy $\pi^{*}$ will be returned to validate and test.

\begin{figure}[htbp] 
    \centering
    \includegraphics[width=1\linewidth]{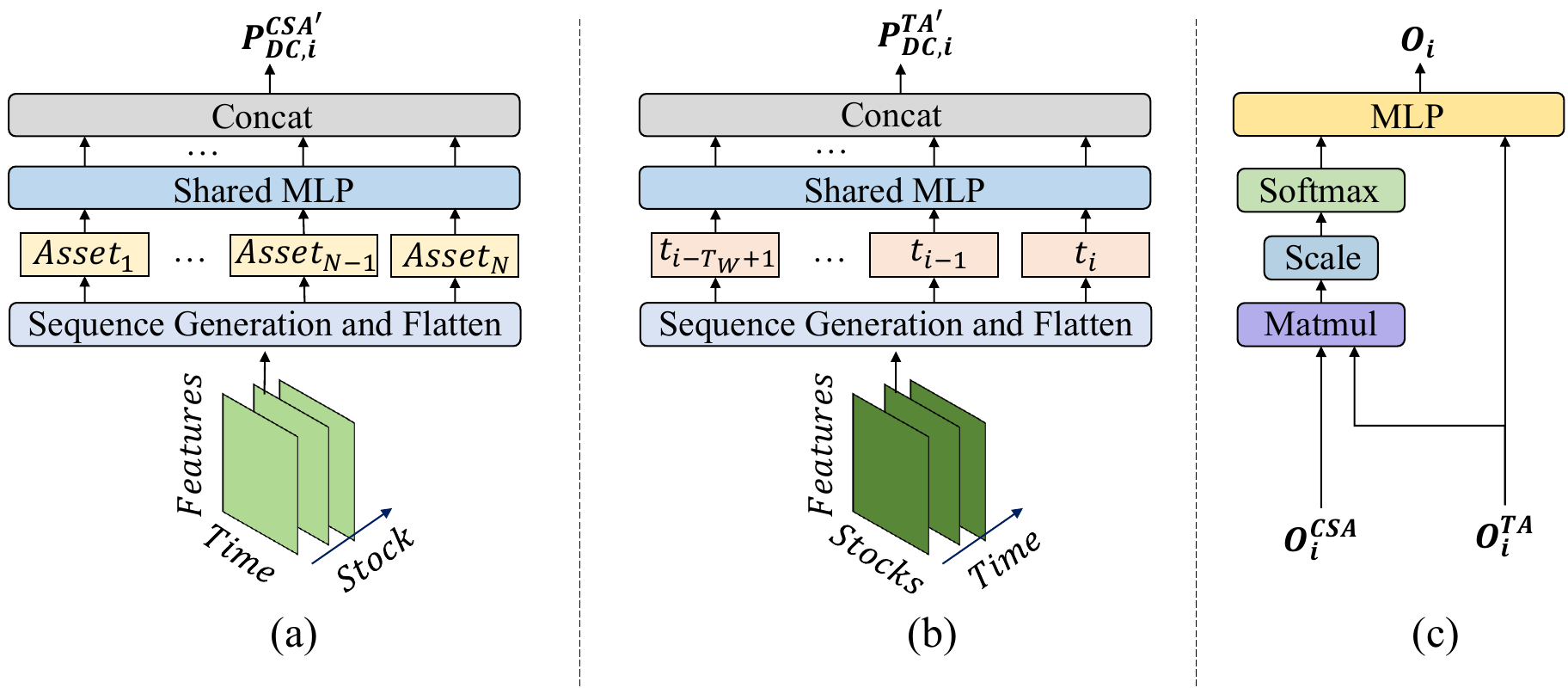}
    \caption{Illustration of the Embedded Block in the CSA and TA Modules, and the Signal Fusion in the Spatial-Temporal Fusion Module}
    \label{fig:component}
\end{figure}

\subsection{Cross-sectional Analysis}

As higher profits usually come with higher risk exposure, diversifying investment risks in PM is a crucial yet challenging task in which trading agents should assign appropriate weights to the assets of different natures for the hedging, thus continuously learning the correlations between assets will facilitate trading agents to better manage risks in such highly turbulent environments. As shown in Fig.~\ref{fig:component}(a), the original DC features will be reshaped and embedded for generating appropriate tokens of sequences before learning the correlations between assets by using the self-attention-based encoders. The output of embedded block of $i^{th}$ agent is denoted as 
    \begin{equation}
        \mathbf{P^{CSA'}_{DC,i}}=f^{CSA}_{i}(\varphi^{CSA}_{i}(\mathbf{P_{DC,i}})), 
        \label{eq:csa}
    \end{equation}

\noindent where $\mathbf{P^{CSA'}_{DC,i}} \in \mathbb{R}^{N \times D}$, $N$ is the length of a sequence (i.e., the number of assets), $ \varphi^{CSA}_{i}(\cdot):\mathbb{R}^{N \times M \times T_{w}}\to  \mathbb{R}^{N \times MT_{w}}$ is the token generation function, and $ f^{CSA}_{i}(\cdot):\mathbb{R}^{N \times MT_{w}}\to  \mathbb{R}^{N \times D}$ is a shared multi-layer perceptron (MLP) function consisting of two linear layers with the $\text{GELU}$ activation function to learn the initial embeddings of each token (i.e., asset) for enhancing feature representation. Afterwards, the attention scores between assets $\mathbf{O^{CSA}_{i}} \in \mathbb{R}^{N \times D} $ can be learnt by the stacked encoders. The optimised attention vector measures the correlation between two different assets in which two assets with similar attention vectors imply more relevant properties in the recent period of time.

\subsection{Temporal Analysis}

In addition to exploring the correlation between two assets, this work tries to investigate the relevance between time points within the given observation period for predicting the price trend at different levels. As depicted in Fig.~\ref{fig:component}(b), unlike the assets as tokens in the CSA module, the temporal analysis regards the time point as the token in a sequence so as to learn the correlations between time points through applying the self-attention-based encoders as well. The output of the embedded block of $i^{th}$ agent can be represented as 
    \begin{equation}
        \mathbf{P^{TA'}_{DC,i}}=f^{TA}_{i}(\varphi^{TA}_{i}(\mathbf{P_{DC,i}})), 
        \label{eq:ta}
    \end{equation}
\noindent where $\mathbf{P^{TA'}_{DC,i}} \in \mathbb{R}^{T_{w} \times D}$, $ \varphi^{TA}_{i}(\cdot):\mathbb{R}^{T_{w} \times N \times M}\to  \mathbb{R}^{T_{w} \times NM}$ is the token generation function, and $ f^{TA}_{i}(\cdot):\mathbb{R}^{T_{w} \times NM}\to  \mathbb{R}^{T_{w} \times D}$ is a shared MLP function with the similar architecture in the CSA module. Besides, with the consideration of chronological effect in a financial market where the impact of price change events will decrease over time, the time sequence mask is introduced in this module. The $\sin(t^{'})$ function is used to generate the mask signals, where the time point $t$ in the observation period is linearly mapped to $t^{'} \in \left [ 0, \frac{\pi}{2}  \right ]$. For instance, the first day of the observation period is assigned to $ t^{'}=0$ with minimum effects while the end day of the observation period is $ t^{'} =\frac{\pi}{2}$ with maximum impacts. Furthermore, by extracting the changes of DC events as high-order DC signals, this module emphasises the features of time points with significant changes. The three kinds of signals will be firstly merged and then fed into the stacked attention-based encoders for optimising the attention scores between two time points $\mathbf{O^{TA}_{i}} \in \mathbb{R}^{T_{w} \times D} $. Accordingly, the underlying trend patterns of two time points are considered to be similar when their attention vectors are close.

\subsection{Ensemble Portfolio Generator}

After collecting the information from the CSA and TA modules, the agents of the proposed framework respectively fuse the attention scores of assets and time points by using the attention mechanism, trying to obtain the attention scores of each asset to every time point within the observation period. As described in Fig.~\ref{fig:component}(c), the output as the suggested portfolio of each agent can be denoted as 
    \begin{equation}
        \mathbf{O}_{i}=\mathbf{V_{i}}(\text{Softmax}(\lambda \mathbf{O^{CSA}_{i}} (\mathbf{O^{TA}_{i}} )^{T})  \mathbf{O^{TA}_{i}})+b_{i}, 
        \label{eq:pogen}
    \end{equation}

\noindent where $\mathbf{O}_{i} \in \mathbb{R}^{N \times 1}$, $\lambda$ is the scale factor, $\mathbf{V_{i}} \in \mathbb{R}^{D \times 1} $ and $b_{i} \in \mathbb{R}^{1 \times 1}$ are the learnable parameters of the MLP.  Ultimately, $\mathbf{O}_{i}$ as the insights of each trading agent in terms of multiple levels of granularity in price changes will be merged to produce a new portfolio for responding to the current financial market. Compared with the portfolio learnt from a single agent, the multiple agents of the proposed framework provide multiple potential portfolios to reduce the probability of biased portfolios according to observing the market features from different ways. This may enhance the capabilities of the proposed framework to handle different financial markets, especially when the market is highly volatile.

\subsection{Policy Optimisation via Reinforcement Learning}

Generally speaking, the PM problem can be modelled as a POMDP and can be optimised by deep RL. Since the reward of suggested portfolios can be immediately collected from the trading environment, the proposed framework can be directly optimised by the policy gradient method rather than using actor-critic RL method~\cite{fujimoto2018addressing} for training a value function to evaluate the actions. In practice, the sum of the logarithm of returns as the reward function of this work is defined as 
    \begin{equation}
        J(\theta)=\frac{1}{T} \log C_{0} \prod_{t=1}^{T} r_{t}, 
        \label{eq:reward}
    \end{equation}
\noindent where $T$ is the number of trading days, $C_{0}$ is the initial portfolio value, $r_{t}$ the rate of daily returns of the executed portfolio. The computed gradient will be backwards propagated to update the CSA, TA, and fusion modules of each agent for continuously learning the trading strategies.

\section{EXPERIMENTS}
\subsection{Experimental Settings}
\noindent \textbf{Datasets:} To demonstrate the effectiveness of the proposed framework in optimising portfolios under the real-world financial environment, three challenging data sets of Dow Jones Industrial Average (DJIA), S\&P 500 and CSI 300 indexes from U.S. and China markets are considered to compare the performance of the MASAAT framework and other representative methods. Furthermore, all the models are trained on ten-year data from January 2008 to December 2017 and then validated by the subsequent data set of three years. Afterwards, the test set from January 2021 to December 2023 is applied to test the validated models of comparative frameworks. It is worth noting that all the involved data sets both include the uptrend period, downtrend period, and some significant global events in order to comprehensively evaluate the models on handling varied market conditions. Besides, the top 10 stocks of each market index are selected as the available stocks in a portfolio according to the company capital. In addition, all the experiments are conducted on a 12-core processor machine with two Nvidia RTX 3090 GPU cards.

\noindent \textbf{Comparative method:} Eight representative methods are selected in this work for the comparison against the proposed MASAAT framework. Constant Rebalanced Portfolio (CRP)~\cite{cover1991universal} is the vanilla method using fixed weights at the whole trading period. Exponential Gradient (EG)~\cite{helmbold1998line} and Passive Aggressive Mean Reversion (PAMR)~\cite{li2012pamr} are the pattern tracking strategies to follow the winner or loser assets in the past period. Moreover, five DL or RL-based approaches are considered. Deep Portfolio Management (DPM)~\cite{jiang2017deep} is based on a convolution neural network while Portfolio Policy Network (PPN)~\cite{zhang2020cost} utilises the LSTM to extract price series information and the convolutional architecture to learn the correlations between stocks. Besides, DeepTrader~\cite{wang2021deeptrader} proposes a graph convolution layer to learn the asset information from markets. Relation-Aware Transformer (RAT)~\cite{xu2021relation} is a transformer-based model for portfolio policy learning. Additionally, this work also includes a well-known actor-critic-based RL approach namely Twin delayed deep deterministic policy gradient (TD3)~\cite{fujimoto2018addressing} for the comparison.

\noindent \textbf{Evaluation Metrics:} Three common performance metrics are selected to compare the concerned approaches in this work. Annualised Return (AR) measures the capability of models to earn profits while Maximum Drawdown (MDD) indicates the risk of maximum losses. Specifically, the Sharpe Ratio (SR) describes the ability of models to balance the profits and risks. The average of each performance indicator over five runs is reported in the following sections. 

\subsection{Performance Analysis}

\begin{table*}[ht]
  \centering
  \caption{The Performance of Representative Approaches Against the Proposed MASAAT Framework on Three Challenging Data Sets} \label{tab:mainres}
    \resizebox{0.9\textwidth}{!}{
    \begin{tabular}{c|ccc|ccc|ccc}
    \hline
    Markets  & \multicolumn{3}{c|}{\textbf{DJIA}} & \multicolumn{3}{c|}{\textbf{S\&P 500}} & \multicolumn{3}{c}{\textbf{CSI 300}} \\
    \hline
    Models & \textbf{AR$(\%)\uparrow$}     & \textbf{MDD$(\%)\downarrow$}     & \textbf{SR$\uparrow$}  & \textbf{AR$(\%)\uparrow$}     & \textbf{MDD$(\%)\downarrow$}     & \textbf{SR$\uparrow$}  & \textbf{AR$(\%)\uparrow$}     & \textbf{MDD$(\%)\downarrow$}     & \textbf{SR$\uparrow$}  \\
    \hline

    CRP        & 13.09 & 18.32 & 0.74 & 19.86 & 22.31 & 0.96 & 0.33   & 25.99 & -0.15 \\
    EG         & 13.02 & 18.35 & 0.74 & 19.90 & 22.29 & 0.96 & 0.30   & 26.08 & -0.15 \\
    PAMR       & 8.82  & \textbf{12.54} & 0.50 & 12.26 & \textbf{14.86} & 0.72 & 4.41   & 24.38 & 0.10  \\
    DPM        & 13.15 & 17.16 & 0.74 & 19.20 & 22.60 & 0.91 & 3.75   & 22.99 & 0.03  \\
    PPN        & 12.03 & 16.77 & 0.74 & 17.99 & 20.78 & 0.93 & 0.65   & 24.11 & -0.14 \\
    DEEPTRADER & 7.81  & 23.21 & 0.34 & 14.01 & 28.70 & 0.57 & -12.90 & 47.78 & -0.76 \\
    RAT        & 12.69 & 17.72 & 0.74 & 19.06 & 21.42 & 0.96 & 0.63   & 24.14 & -0.14 \\
    TD3        & 13.10 & 18.31 & 0.74 & 19.86 & 22.31 & 0.96 & 0.34   & 26.00 & -0.15 \\
    \textbf{MASAAT}     & \textbf{14.28} & 16.24 & \textbf{0.81} & \textbf{21.57} & 19.84 & \textbf{1.03} & \textbf{5.13}   & \textbf{21.16} & \textbf{0.11}  \\
    \hline
    \end{tabular}%
    }
\end{table*}

\noindent \textbf{Results on DJIA:} Table~\ref{tab:mainres} demonstrates the performance of all comparative methods in the DJIA market index. The proposed MASAAT framework achieves the highest AR of 14.28\% during the backtesting period of three years, which is at least 1.13\% higher than that of the best baseline frameworks achieved by the DPM. Even though the PAMR may suffer the lowest loss of 12.54\% MDD, it earns very low profits and SR due to the conservative trading strategies. Compared with other methods, the MASAAT manages relatively low risks when pursuing higher returns, thus achieving the highest SR over 0.8 which demonstrates the capability of the proposed framework to balance the overall returns and investment risks.

\noindent \textbf{Results on S\&P 500:} With the comparison of DL or RL-based approaches, the performance of the trading strategies proposed by the MASAAT has an impressive enhancement in all the three metrics. Particularly, the MASAAT increases the AR by 2\% and reduces the MDD risks by 2\% when compared with the RAT framework in which the attention-based mechanism is adopted to investigate the correlations between stocks. Similarly, the PAMR demonstrates good risk management in the S\&P 500 index, but its AR drops by around 10\% than that of the MASAAT as the PAMR may follow the asset with lowest returns during most of the trading period.

\noindent \textbf{Results on CSI300:} From the reported results in Table~\ref{tab:mainres}, the AR of the most baseline methods is worse than the reference zero-risk asset in which the yield of treasury bond in China is around 3\% per year, which may lead to a negative SR in the most of baseline approaches. In such volatile markets, by possibly executing conservative trading strategies, the PAMR can obtain relatively high profits, achieving an AR of 4.41\% and an SR of 0.1. Similar to the performance achieved in the DJIA and S\&P 500 indexes, the proposed MASAAT framework outperforms all the comparative approaches in all metrics in terms of profits and risks. 

\begin{figure}[htbp]
  \centering
  \includegraphics[width=0.85\linewidth]{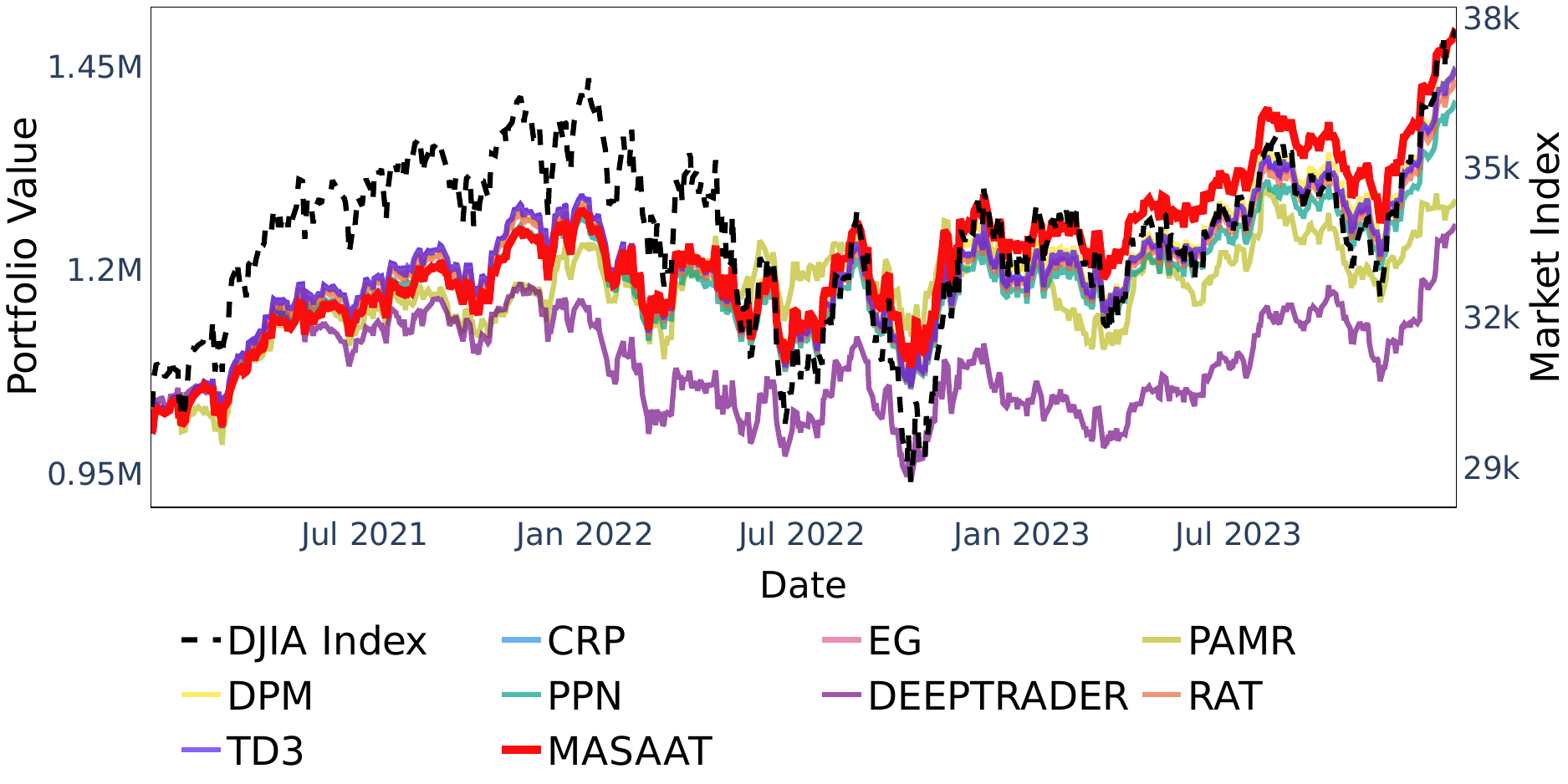}
  \caption{The Portfolio Values of Different Approaches on the DJIA Index}
  \label{fig:pvtrenddjia}
\end{figure}

\begin{figure}[htbp]
  \centering
  \includegraphics[width=0.85\linewidth]{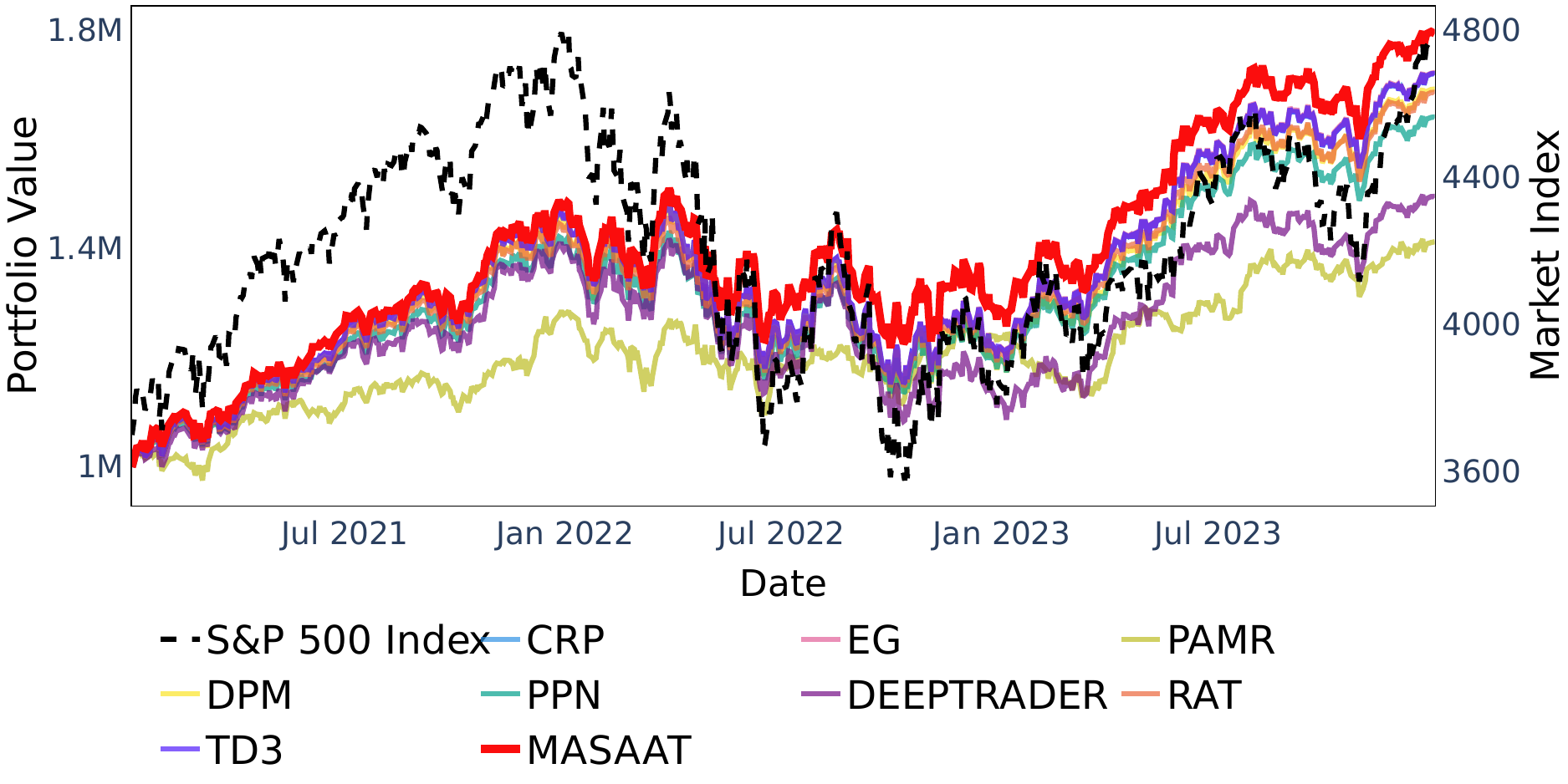}
  \caption{The Portfolio Values of Different Approaches on the S\&P 500 Index}
  \label{fig:pvtrendsp500}
\end{figure}

\begin{figure}[htbp]
  \centering
  \includegraphics[width=0.85\linewidth]{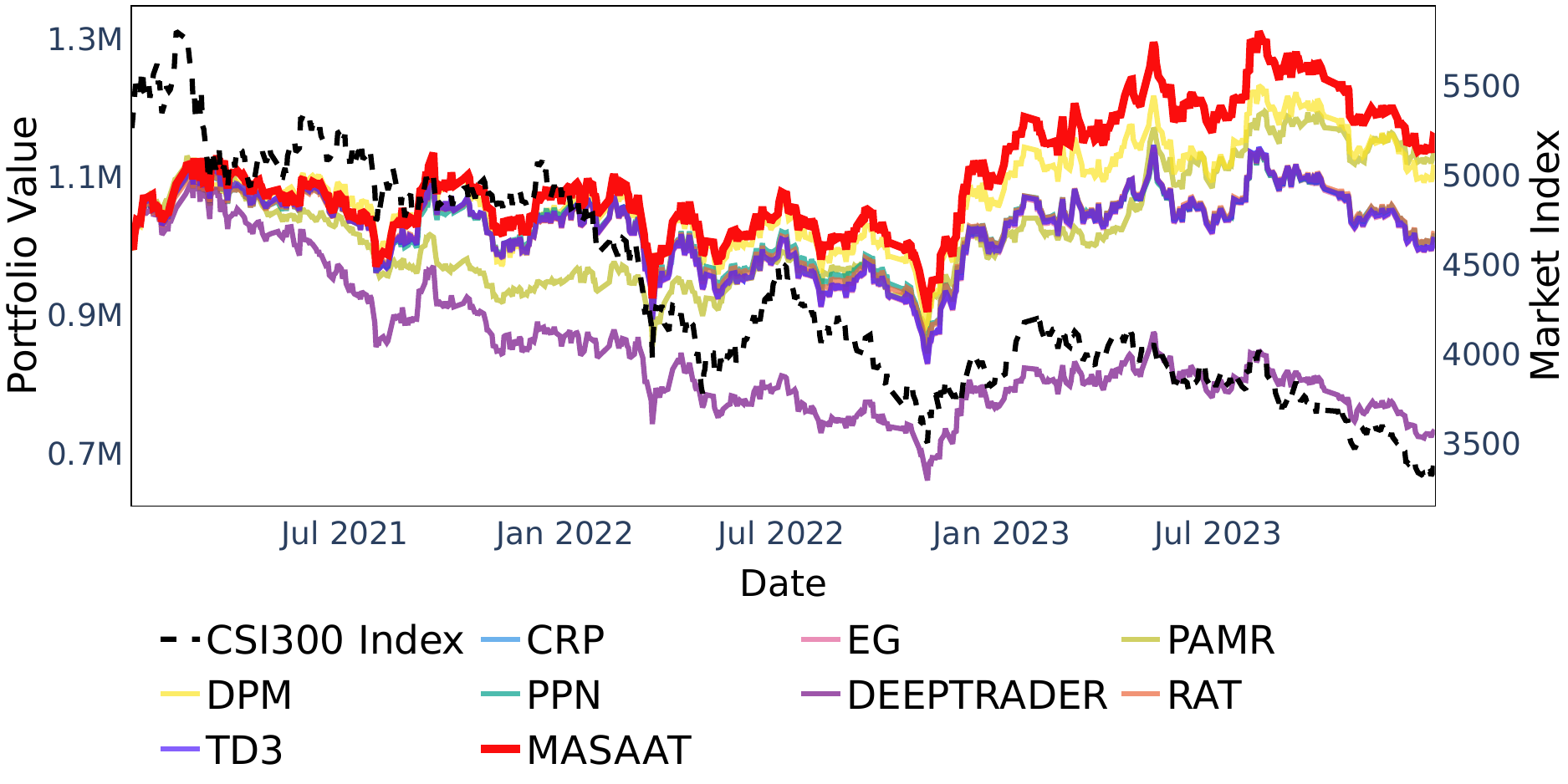}
  \caption{The Portfolio Values of Different Approaches on the CSI 300 Index}
  \label{fig:pvtrendcsi300}
\end{figure}

Fig.~\ref{fig:pvtrenddjia} to Fig.~\ref{fig:pvtrendcsi300} shows the trends of portfolio values of all comparative approaches and market index at the whole test period from 2021 to 2023 on the three market indexes. The MASAAT attains remarkable performance on different market conditions such as the upward period, downward period, and fluctuating period. Especially when the market indexes dramatically drops in 2022, the proposed MASAAT can well manage the risks to avoid huge losses during the period. All in all, integrating multiple agents to observe the changes of assets at different levels can effectively enhance the capability of the MASAAT framework to adapt to the financial environments under varied market conditions.

\noindent \textbf{Ablation Study and Key Parameter Analysis}

\begin{table}[ht]
  \centering
  \caption{The Ablation Study and Key Parameter Analysis of the Proposed MASAAT Framework on the DJIA Index} 
  \label{tab:abla}
    \begin{tabular}{c|ccc}
    \hline
    Models & \textbf{AR$(\%)\uparrow$}     & \textbf{MDD$(\%)\downarrow$}     & \textbf{SR$\uparrow$}  \\
    \hline
    MASAAT-w/o TS & 13.04 & 17.84 & 0.74 \\
    MASAAT-w/o DC & 13.09 & 18.30 & 0.74 \\
    MASAAT-1    & 13.11 & 18.31 & 0.74 \\
    MASAAT-3    & 14.28 & 16.24 & 0.81 \\
    MASAAT-5    & 13.56 & 18.13 & 0.76 \\

    \hline
    \end{tabular}%
\end{table}

Table~\ref{tab:abla} reviews the compared results of the ablation study and key parameter analysis of the proposed MASAAT framework on the DJIA index. The MASAAT-w/o TS denotes the MASAAT framework removing the trading agent that directly deals with the conventional time-based price series while the MASAAT-w/o DC does not consider the DC-related data. Compared with the original MASAAT framework learning and analysing from both the price series and DC features, the framework considering either time-based price series or DC features may not effectively capture the trend information from the current market and possibly produce biased trading decisions, thus the AR drops to around 13\% and the maximum loss increases by 2\%. In addition, this work investigates the impact of the number of DC-oriented agents involved in the proposed MASAAT framework. The MASAAT-1/3/5 are the MASAAT framework integrated with 1, 3, or 5 agent(s) to deal with the DC features generated by the corresponding number of DC thresholds. It should be noted that the reported results of the MASAAT in Table~\ref{tab:mainres} are based on three DC-oriented trading agents to extract the DC features from three different DC thresholds.  From the results described in Table~\ref{tab:abla}, by increasing the number of trading agents, the AR rises from 13.09\% at the MASAAT-w/o DC to 14.28\% at the MASAAT-3 and subsequently drops to 13.56\% when the number of DC-oriented agents is set to five in which the MASAAT framework can observe the trend patterns from the higher DC thresholds. However, the higher DC threshold may introduce the DC signals that stay at the same value in most of the trading period as the significant price changes do not frequently occur, which may not help the MASAAT framework analyse the assets and possibly generate biased decisions.

\section{CONCLUDING REMARKS}

Financial Portfolio optimisation has been studied for a few decades yet is still a very challenging and significant task for investors to balance investment returns and risks under different financial market conditions. There are many studies trying to use various deep or reinforcement learning approaches such as convolution-based, recurrent-based, and graph-based neural networks to capture the spatial and temporal information of assets in a portfolio.  However, due to the conventional price series involving a lot of noise, the trend patterns may not be easy to discover by most of the existing methods under the highly turbulent financial market. In this work, a multi-agent and self-adaptive portfolio optimisation framework integrated with attention mechanisms and time series namely the MASAAT is proposed in which multiple trading agents are introduced to analyse price data from various perspectives to help reduce the biased trading actions. In addition to the conventional price series, the directional changes-based data are considered to record the significant price changes in different levels of granularity for filtering any plausible noise in financial markets. Furthermore, the attention-based cross-sectional analysis and temporal analysis in each agent are adopted to capture the correlations between assets and time points within the observation period in terms of different viewpoints, followed by a spatial-temporal fusion module attempting to fuse the learnt information. Lastly, the portfolios suggested by all agents will be further merged to produce a newly ensemble portfolio so as to quickly respond to the current financial environment. The empirical results on three challenging data sets of DJIA, S\&P 500, and CSI 300 market indexes reveal the strong capability of the proposed MASAAT framework to balance the overall returns and portfolios risks against the state-of-the-art approaches.

In the future, the proposed MASAAT framework sheds lights on many interesting directions for further investigation. First, the trading signals provided by different trading agents can be fused by some intelligent methods. For example, the extension can assign different weights to the signals generated by agents according to their historical performance. Besides, the proposed framework can absorb multimodal information through creating a new agent to learn the sentiment information from current markets. Lastly, the MASAAT framework can be further applied to various financial applications such as the high-frequency trading in the cryptocurrency market, foreign exchange trading, option dealing, etc.

\bibliographystyle{IEEEtran}
\bibliography{MASAAT}

\end{document}